# Gaussian and sinc shaped few-cycle pulse driven ultrafast coherent population transfer in $\Lambda$ - like atomic systems

**Parvendra Kumar and Amarendra K. Sarma***
Department of Physics, Indian Institute of Technology Guwahati, Guwahati-781039, Assam, India.
*Electronic address: aksarma@iitg.ernet.in

We report and propose a simple scheme to achieve efficient and fast coherent population transfer (CPT) by utilizing either a nonlinearly chirped Gaussian shaped few-cycle laser pulse or an unchirped sinc-shaped few-cycle laser pulse. The proposed scheme is shown to be fairly robust against the variation of the laser parameters such as temporal pulse width, chirp rates, carrier envelope phases and Rabi frequencies. We find that compared to the so-called stimulated Raman adiabatic passage (STIRAP) technique, our scheme for complete population transfer with few-cycle Gaussian shaped laser pulses requires less pulse area.

## I. INTRODUCTION

Coherent control of complete population transfer from an initial quantum state to a desired quantum state and creation of arbitrary coherence between two quantum states has been a major theme of atomic and molecular physics for quite some time now [1]. In particular, coherent control of population transfer is necessary for many well-known applications, including collision-dynamics, atomic interferometry, spectroscopy and optical control of chemical reactions etc. Coherent population transfer (CPT) has found its relevance even in nuclear physics [2]. For robust and efficient controlling of population transfer between quantum states, many novel strategies have been proposed and exploited by several authors. Stimulated Raman adiabatic passage (STIRAP) [3-6], adiabatic rapid passage (ARP) [7], Raman chirped adiabatic passage (RCAP) [8-10] and temporal coherent control (TCC) [11-12] etc. are some of the well-known methods. Recently, the field of coherent control of atoms and molecules have received tremendous boost owing to the recent progress in the generation of the femtosecond and attosecond laser pulses and its possible future applications [13-18]. For example, quantum coherent control of physical and chemical processes and also of attosecond electronic dynamics by use of frequency- and amplitude-chirped few-cycle pulses is reported by some authors [19-20]. In the context of coherent population transfer (CPT), the use of these so-called ultrashort laser pulses has certain advantages, such as, easy access to first electronic states of many molecules and extremely fast transfer process which may be completed on a time scale much shorter than the typical time between collisions of atoms or molecules [21]. Cheng and Zhou [22] demonstrated numerically, without using the so-called rotating wave approximation (RWA), the ultrafast population transfer in $\Lambda-$ like three state atomic systems with frequency chirped few-cycle femtosecond laser pulses, though the scheme was not found to be robust against the variation of laser pulse parameters. In passing, it should be mentioned that RWA may not hold when one deal with few-cycle pulse related phenomena and should work in the non-RWA regime [13, 23-26]. Recently, E. A. Shapiro et al. [27-28] proposed a new method, the so called piecewise rapid adiabatic passage (PAP), for executing complete population transfer between quantum states in a piecewise manner using a series of femtosecond laser pulses. The proposed method is found to be robust against the variation in the absolute and relative intensities, durations, and time ordering of the pulses. More recently, X. H. Yang [24] exploited the STIRAP method to demonstrate the sufficiently robust and complete population transfer between the quantum states in $\Lambda-$like three state atomic systems driven by few-cycle femtosecond laser pulses. However, for complete population transfer with STIRAP technique, the time separated but partially-overlapping pump and Stokes pulses should be applied in counterintuitive order i.e. there should be complete control over the time lag between the pump and Stokes pulses. It is well known that the STIRAP technique is successfully implemented with continuous or narrow-band laser pulses. However, as discussed by many authors, when one needs to deal with few-cycle pulses or broad-band pulses, this method may not be a suitable one [21,29-30]. Moreover, for complete population transfer between the quantum states, the STIRAP and rapid adiabatic passage are generally energetically expensive, for example, relative to a $\pi-$pulse technique [31-32]. On the other hand, sometimes the $\pi-$pulse technique is not robust against the variation of the laser pulse parameters. In this work, we propose a relatively simple scheme, in which no time separation is required between the pump and Stokes pulses. Our method is energetically efficient and ultrafast. We show near complete CPT in the given $\Lambda-$like three state atomic system which is driven by two simultaneously interacting nonlinearly chirped Gaussian shaped few-cycle laser pulses or by two sinc shaped unchirped few-cycle laser pulses, subject to the judicious choice of laser pulse parameters. It may be mentioned that shaped pulses are also studied to enhance transient population of excited states [26] and obtain total inversion of electronic state population in molecules [33]. Again, it may be noted that nonlinearly chirped laser pulses have been found suitable for many applications [34-38]. In this work, the phenomenon of coherent population transfer is investigated by numerically solving the appropriate density matrix equations beyond the rotating wave approximation. In Sec. II we present the optical Bloch equations that describe the interaction of the $\Lambda$ system with the few-cycle laser pulses. Sec. III contains our simulated results and discussions followed by conclusions in Sec. IV.

## II. THE MODEL

Our analysis is based on the scheme depicted in Fig.1. We consider a $\Lambda$-like atomic system interacting with two few-cycle laser pulses. The electric field of the linearly polarized laser interacting between $|3\rangle$ and $|1\rangle$ is given by $\vec{E}_1 = \vec{E}_{10} f(t) \cos(\omega_{10} t + \delta_1(t))$, where $\vec{E}_{10}, f(t)$ and $\omega_{10}$ are respectively, the amplitude, the field envelope and the carrier frequency of the pulse. Exactly analogous expression for the linearly polarized laser pulse interacting between $|3\rangle$ and $|2\rangle$ is given by $\vec{E}_2 = \vec{E}_{20} f(t) \cos(\omega_{20} t + \delta_2(t))$. For the Gaussian shaped few-cycle laser fields, $f(t) = \exp[-(t/\tau)^2]$, $\delta_1(t) = \chi_1 t^3$ and $\delta_2(t) = \chi_2 t^3$. Here $\chi_1$ and $\chi_2$ are the respective chirp rate of the Gaussian pulses. On the other hand, for the sinc-shaped few-cycle pulses, $f(t) = \sin(t/\tau)/(t/\tau)$ and $\delta_1 = \phi_1$ and $\delta_2 = \phi_2$, where $\phi_1$ and $\phi_2$ are the respective carrier-envelope phases. $\tau_p$ is the temporal width of the laser pulses. For the Gaussian pulse, $\tau_p = 1.177 \tau$ while $\tau_p = 2.783 \tau$ for the sinc pulse.

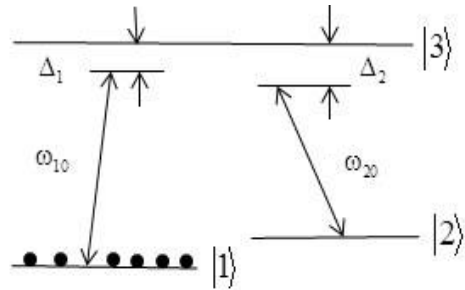

Fig. 1 Schematic of $\Lambda$-like system with two acting few-cycle laser pulse

In the given scheme, we assume that only $|3\rangle \to |1\rangle$ and $|3\rangle \to |2\rangle$ transitions are dipole allowed while $|2\rangle \to |1\rangle$ transitions are forbidden. The Hamiltonian of the system is given by $\hat{H} = \hat{H}_0 + \hat{H}_{int}$ where $\hat{H}_0 = \hbar\omega_1 |1\rangle\langle 1| + \hbar\omega_2 |2\rangle\langle 2| + \hbar\omega_3 |3\rangle\langle 3|$ and $\hat{H}_{int} = -\vec{\mu}.\vec{E} = -\hbar\Omega_{31}(t)|3\rangle\langle 1| - \hbar\Omega_{32}(t)|3\rangle\langle 2| + \text{h.c.}$
Here $\Omega_{31}(t) = \mu_{31}\vec{E}_1(t)/\hbar$ and $\Omega_{32}(t) = \mu_{32}\vec{E}_2(t)/\hbar$ are the time dependent Rabi frequency for the transition with electric dipole moment $\mu_{31}$ and $\mu_{32}$ respectively. The Bloch equations, without invoking the so called rotating wave approximation, describing the temporal evolution of the density matrix elements are:

$$\dot{\rho}_{31} = -i\omega_{31}\rho_{31} + i\Omega_{32}(t)\rho_{21} - i\Omega_{31}(t)(\rho_{33} - \rho_{11})$$
$$\dot{\rho}_{32} = -i\omega_{32}\rho_{32} + i\Omega_{31}(t)\rho_{12} - i\Omega_{32}(t)(\rho_{33} - \rho_{11})$$
$$\dot{\rho}_{21} = -i\omega_{21}\rho_{21} + i\Omega_{32}(t)\rho_{31} - i\Omega_{31}(t)\rho_{23}$$
$$\dot{\rho}_{11} = i\Omega_{31}(t)(\rho_{31} - \rho_{13})$$
$$\dot{\rho}_{22} = i\Omega_{32}(t)(\rho_{32} - \rho_{23})$$
$$\dot{\rho}_{33} = i\Omega_{31}(t)(\rho_{13} - \rho_{31}) + i\Omega_{32}(t)(\rho_{23} - \rho_{32})$$

(1)

Here $\omega_{ij} = \omega_i - \omega_j$. It may be noted that $\rho_{ij} = \rho_{ji}^*$.

### III. RESULTS AND DISCUSSIONS

We solve Eq. (1) numerically using a standard fourth-order Runge-Kutta method. We assume that initially all the atoms are in the ground state $|1\rangle$. We use the following typical parameters: $\omega_{31} = \omega_{10} = 3.0$ rad/fs, $\omega_{21} = 0.4$ rad/fs, $\omega_{32} = \omega_{20} = 2.6$ rad/fs, $\Omega_{31} = 0.76$ rad/fs, $\Omega_{32} = 0.79$ rad/fs, $\chi_1 = \chi_2 = 0.016$ fs$^{-3}$, $\phi_1 = \phi_2 = 0$. The temporal pulse width is taken to be, $\tau_p = 4.70$ fs and 5.06 fs respectively for the Gaussian and the sinc pulses. For our chosen parameters, the pulse areas are too small for the so-called adiabatic condition to be fulfilled. It is worthwhile to note that in the usual adiabatic passage scheme for population transfer between the initial state and the final one, the adiabatic condition can be written as: $\Omega\tau \gg \pi$, where $\Omega = \Omega_{31}(0) = \Omega_{32}(0)$ is the maximal Rabi frequency [39]. In practical applications the pulse area should exceed $10\pi$, i.e. $\Omega\tau > 10\pi$ to provide efficient population transfer via the adiabatic passage scheme [40]. We find that for complete population transfer, with our proposed scheme, the total temporal area of the Gaussian pulses is calculated to be $3.49\pi$, while it is $13.53\pi$ for the STIRAP scheme with same laser pulse parameters. Fig. 2 and Fig. 3 depict the respective temporal evolution of the populations $\rho_{11}, \rho_{22}$ and $\rho_{33}$ when chirped Gaussian pulses and unchirped sinc-pulses are used.

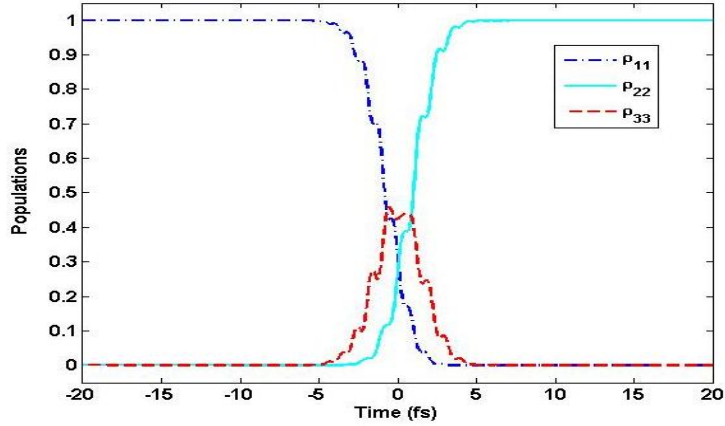

Fig. 2 (Color online) Temporal evolution of populations with the nonlinearly-chirped Gaussian shaped few-cycle pulse

It is clear from Fig. 2 that one can obtain complete population transfer **(99.94 %)** from the ground state |1> to the state |2> using two nonlinearly chirped Gaussian shaped laser pulses. On the other hand, as evident from Fig. 3, near complete population transfer **(99.05 %)** from |1> to |2> is possible due to the simultaneous interaction of two sinc-shaped pulses with the three-level atomic system.

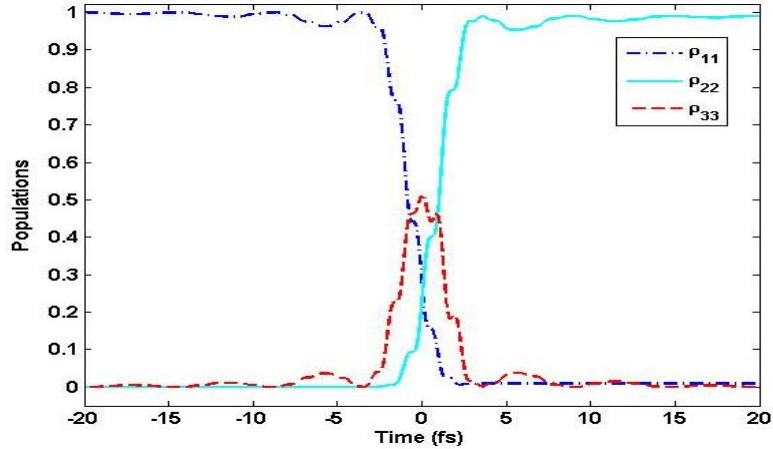

Fig. 3 (Color online) Temporal dynamics of populations with the Sinc-shaped few-cycle pulse

These results could be explained on the basis of the so-called Stimulated Emission Pumping (SEP) [31]. In SEP with continuous laser or short laser pulses, all the relaxation processes in atomic systems takes place on the time scale shorter than the interaction time. Hence the maximum amount of population transfer between the quantum states is restricted by the spontaneous emission. For example, one can achieve maximum 30 % population transfer in $\Lambda$ – like three level atomic systems with SEP technique [31]. However, in our scheme, we have shown that almost complete population transfer in $\Lambda$ – like atomic systems is possible owing to the use of few-cycle pulses where interaction takes on a time scale shorter than that of the relaxation processes. Non-adiabatic consequences on the temporal evolution of the populations could be observed from Fig. 2 and 3. Unlike the adiabatic passage techniques, population in quantum state $|3\rangle$ during the intermediate time is approaching a large value ( $\rho_{33}$ =45 % for Gaussian pulse and $\rho_{33}$ = 46 % for sinc pulse) as the adiabatic criteria is not fulfilled for the chosen laser pulse areas. However, finally the quantum state $|2\rangle$ receives almost all the populations in both cases. The non-RWA effects on the temporal evolution of the populations could also be observed from Fig. 2 and 3. Some authors have pointed out that when the few-cycle laser pulses are considered, the time-derivative driven nonlinearities will have a significant impact on the interaction of the laser pulses with the atomic medium which may lead to strong oscillation features during the evolution of the populations [41-42]. These features are not present in the RWA solutions. Now, in order to have some insight or understanding why the use of Gaussian shaped or sinc-shaped few-cycle pulse results in almost similar behaviour with regard to population transfer, in Fig. 4 we plot the temporal evolution of both the pulses.

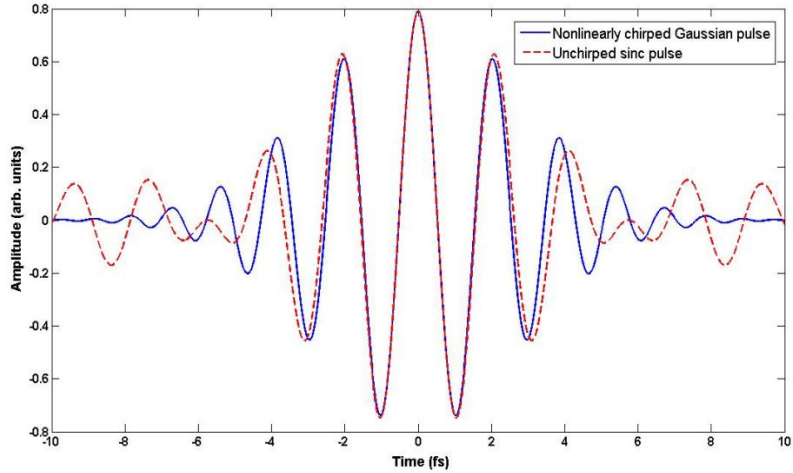

Fig. 4 (Color online) Temporal evolution of nonlinearly-chirped Gaussian and unchirped sinc shaped pulses.

Fig. 4 reveals that the nonlinearly-chirped Gaussian and the unchirped sinc shaped pulses are equivalent to each other, particularly in the temporal range from -2 to +2 fs. For other temporal range from -6 to -2 fs and +2 to +6 fs, the amplitude of the nonlinearly chirped Gaussian pulse is slightly greater than that of the unchirped sinc shaped pulse. Also, the carrier oscillation frequency of the Gaussian shaped few-cycle pulse is slightly larger, owing to the nonlinear chirp, than that of the unchirped sinc shaped few-cycle pulse in that temporal range. This might be the reason behind the almost similar nature of interaction of the pulses with the atomic system. Hence, we may conclude that, subject to the chosen parameters, the nonlinearly chirped Gaussian and the unchirped sinc shaped few-cycle pulses exhibit almost identical behaviour. It may be observed from Fig. 4 that one ($\Delta_1 = \omega_{31} - \omega_{10} = \Delta_2 = \omega_{32} - \omega_{20} = 0$) and two-photon ($\Delta = \Delta_1 - \Delta_2 = 0$) resonance conditions are fulfilled for unchirped sinc-shaped laser pulse. However, for the nonlinearly chirped Gaussian laser pulse, one photon resonance condition is partially fulfilled during the intermediate time of interaction while the two-photon resonance condition is fulfilled during the whole interaction. It is important to verify the robustness of the scheme against the variation of the chirp rate and the temporal pulse width of the Gaussian shaped few-cycle pulse. So, in Fig. 5 we present the simulation result for the variation of the final population transfer to the state |2>, i.e. $\rho_{22}(\infty)$ with $\tau_p$, while in Fig. 6 we check the robustness of the scheme against chirp rates for the nonlinearly chirped Gaussian shaped few-cycle pulses.

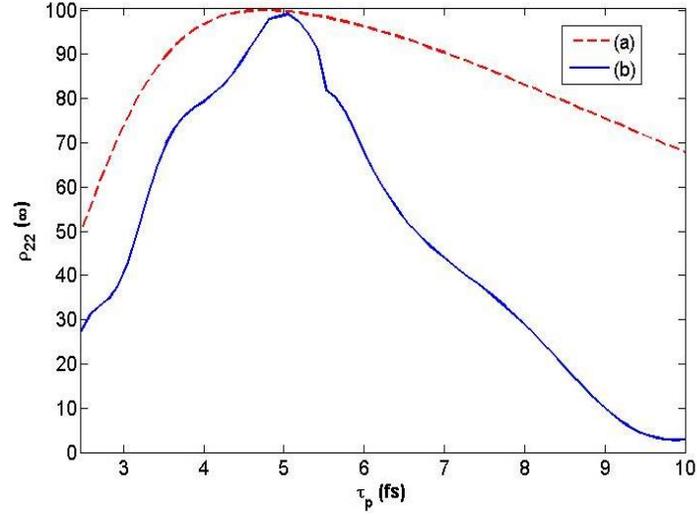

Fig. 5 (Color online) Final population to quantum state |2> as a function of (a) Temporal pulse width $\tau_p$ of the nonlinearly chirped Gaussian pulse and (b) temporal pulse width $\tau_p$ of the sinc pulse. Here all the other parameters are kept constant.

It can be seen from Fig. 5(a) that the final population transfer $\rho_{22}(\infty)$ to the quantum state $|2\rangle$ is sufficiently robust against the variation (4-6 fs) of temporal width of Gaussian pulse, while Fig. 5(b) shows that the final population transfer $\rho_{22}(\infty)$ to quantum state $|2\rangle$ is robust against the small variation (4.94-5.17 fs) of temporal width of the sinc pulse.

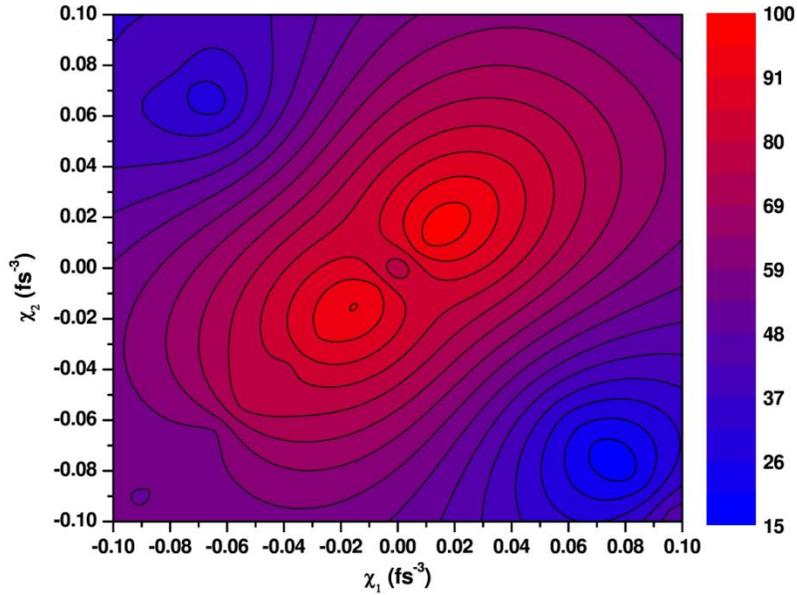

Fig. 6 (Color online) Final population transfer $\rho_{22}(\infty)$ to quantum state |2> as a function of chirp rates, $\chi_1$ and $\chi_2$ and other parameters are same as those in Fig. 2.

It could be observed from Fig. 6 that the final population transfer $\rho_{22}(\infty)$ is sufficiently robust against the variation of the chirp rates, $\chi_1$ and $\chi_2$. It can be seen that the final population transfer $\rho_{22}(\infty)$ to the quantum state $|2\rangle$ is robust against the small variation (0.012-0.020 fs$^{-3}$) of the chirp rates $\chi_1$ and $\chi_2$.

In Fig. 7 we plot the variation of the final population $\rho_{22}(\infty)$ with the carrier envelope phases to check the robustness of the scheme for sinc-shaped few-cycle pulses.

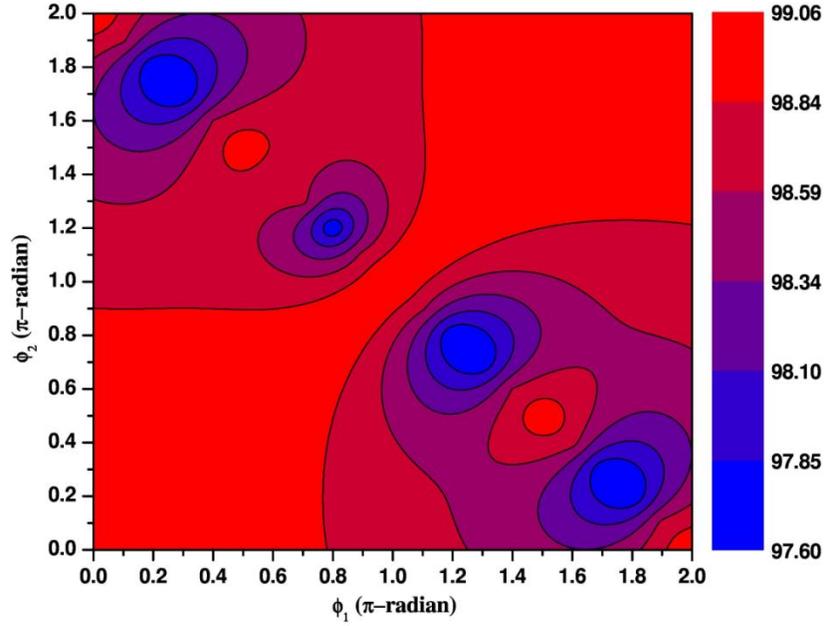

Fig. 7 (Color online) Final population transfer $\rho_{22}(\infty)$ to quantum state |2> as a function of (a) carrier-envelope phases, $\phi_1$ and $\phi_2$ in the unit of $\pi-$ radian and other parameters are same as those in Fig. 3.

We find that, from Fig. 7, the final population transfer $\rho_{22}(\infty)$ is highly robust against the variation of the carrier-envelope phases, $\phi_1$ and $\phi_2$ of the sinc-shaped few cycle pulses.

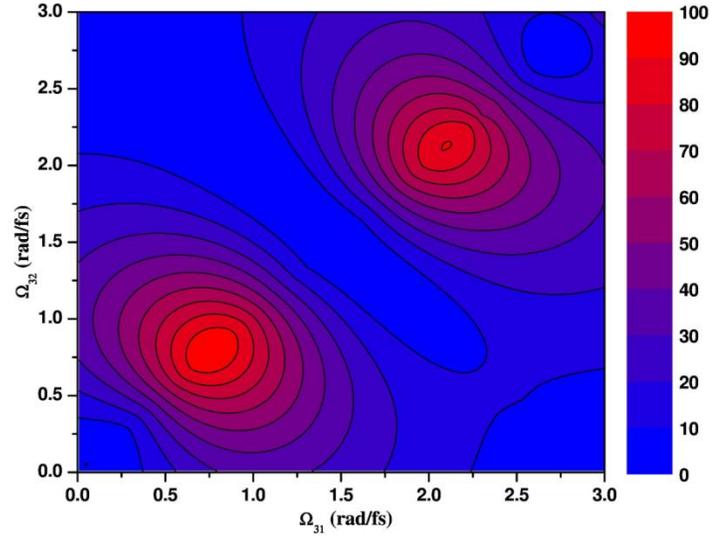

Fig. 8 (Color online): Contour maps of the final population (in %) transfer for varying Rabi frequencies $\Omega_{31}$ and $\Omega_{32}$ of nonlinearly chirped Gaussian pulse and other parameters are same as those in figure 2.

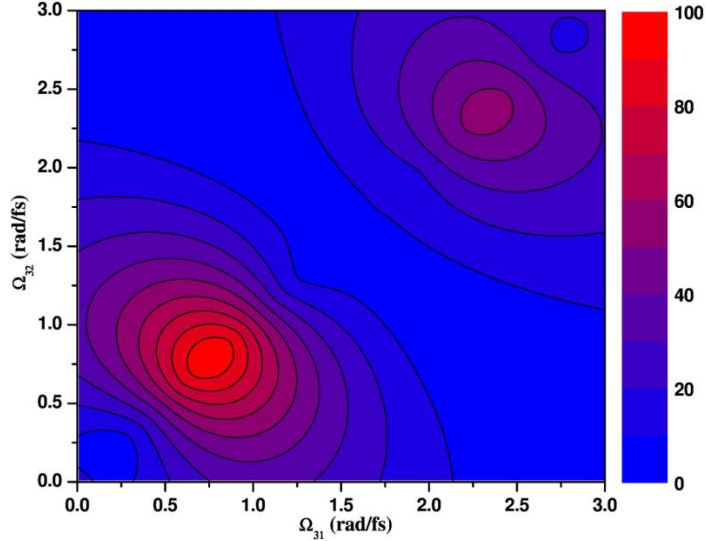

Fig. 9 (Color online) Contour maps of the final population (in %) transfer for varying Rabi frequencies $\Omega_{31}$ and $\Omega_{32}$ of unchirped sinc pulse and other parameters are same as that in figure 3.

Finally, we test the robustness of our proposed scheme against variation of the Rabi frequencies of the few-cycle laser pulses considered in this work. In Fig. 8, we depict the contour map of the final population to the quantum state |2>, i.e. $\rho_{22}(\infty)$, against Rabi frequencies $\Omega_{31}$ and $\Omega_{32}$ of the nonlinearly chirped Gaussian shaped few-cycle pulse. A careful inspection of Fig. 8 reveals that the final population transfer is fairly robust against the variation of the Rabi frequencies $\Omega_{31}$ and $\Omega_{32}$ in the range of 0.70-0.92 rad/fs, which amounts to more than 95 percent population. Population in the range of 88-90% could be possible for variation of the Rabi frequencies in the range 2.30-2.40 rad/fs. A similar contour map for the sinc shaped few-cycle pulse is depicted in Fig. 9. The population transfer to the quantum state |2> exhibits sufficient robustness with the small variation of Rabi frequencies. In fact one can obtain more than 95% population transfer for the variation of $\Omega_{31}$ and $\Omega_{32}$ in the range of 0.70-0.85 rad/fs.

## IV. CONCLUSIONS

We have demonstrated almost complete population transfer to the target quantum state with nonlinearly chirped Gaussian shaped and unchirped sinc-shaped few-cycle laser pulses even when the adiabatic condition is not fulfilled. Population transfer with nonlinearly chirped Gaussian pulse is found to be sufficiently robust against the variation of temporal pulse width, Rabi frequencies and the chirp rates. However, the population transfer with unchirped sinc-shaped pulse is found to be highly robust against the variation of carrier-envelope phase and fairly robust against the variation of the temporal pulse width and Rabi frequencies. We find that compared to the so-called STIRAP technique, our scheme for complete population transfer with few-cycle Gaussian shaped laser pulses require less pulse area. Hence, nonlinearly chirped Gaussian shaped few-cycle laser pulses or the sinc-shaped few cycle laser pulses, even with highly fluctuating carrier-envelope phase, may be employed for an efficient and ultrafast coherent population transfer with judicious choice of laser pulse parameters.